\documentclass{article}
\usepackage{amssymb}
\usepackage{amsfonts}
\usepackage{amsmath}

\input{tcilatex}
\begin{document}

\title{On the Weyl Tensor for a Curved Spacetime Endowed with a Torsionful
Affinity}
\author{J. G. Cardoso\thanks{%
jorge.cardoso@udesc.br} \\
Department of Mathematics\\
Centre for Technological Sciences-UDESC\\
Joinville 89223-100 SC\\
Brazil.}
\maketitle
\date{ }

\begin{abstract}
We transcribe into the framework of the torsionful version of the $%
\varepsilon $-formalism of Infeld and van der Waerden the world definition
of the Weyl tensor for a curved spacetime that occurs in the realm of
Einstein-Cartan's theory. The resulting expression shows us that it is not
possible to attain any general condition for conformal flatness in such a
spacetime even if wave functions for gravitons are eventually taken to
vanish identically. A short discussion on the situation concerning the
limiting case of general relativity is presented thereafter.
\end{abstract}

\section{Introduction}

The most important physical significance of the Weyl tensor for a curved
spacetime having torsionlessness [1,2] relies upon the fact that its
two-component spinor expression, as provided by either of the $\gamma
\varepsilon $-formalisms of Infeld and van der Waerden [3], carries locally
only suitable couplings between metric spinors and wave functions for
gravitons [2]. Such wave functions thus show up in both formalisms as
totally symmetric parts of certain curvature spinors, being at every
spacetime point likewise taken to describe microscopically the degrees of
freedom of a metric tensor while vanishing identically in the case of a
conformally flat spacetime. In this torsionless context, the vanishing of a
Weyl tensor constitutes the definitive condition for conformal flatness.

In the present paper, we first transcribe into the framework of the
torsionful version of the $\varepsilon $-formalism [4] the world definition
of the Weyl tensor for a curved spacetime equipped with a torsionful
affinity. The spinor expression which results from our transcription shows
us that it is not possible to attain any general condition for conformal
flatness in such a spacetime even if wave functions for gravitons are
eventually taken to vanish identically. A short discussion on the situation
concerning the limiting case of general relativity is then given. For the
sake of consistency, we will have to call for some of the world and spinor
structures that take place in the context of Einstein-Cartan's theory such
as exhibited in Refs. [4,5]. It shall become quite clear that all the spinor
expressions arising herein still remain formally valid when the pertinent
procedures are occasionally shifted to the framework of the torsional $%
\gamma $-formalism.

The principal motivation for elaborating upon the situation being
considered, stems from the belief that our work might provide some original
material which can be of some importance as far as the cosmological models
based upon Einstein-Cartan's theory are concerned [6-8]. In Section 2, we
bring out the structures needed for working out the transcription procedures
consistently (Section 3). The limiting case is considered in Section 4 along
with some physical remarks. The symbolic configurations displayed in the
work of Ref. [4] shall be utilized many times in what follows, but we will
refer to them explicitly just a few times. All the conventions adhered to in
Ref. [5] will be taken for granted at the outset.

\section{Some Basic Structures}

Let us consider a curved spacetime endowed with a symmetric metric tensor $%
g_{\mu \nu }$ of signature $(+---)$ together with a torsionful covariant
derivative operator $\nabla _{\mu }$ that satisfies the metric compatibility
condition $\nabla _{\mu }g_{\lambda \sigma }=0.$ The Weyl tensor of $\nabla
_{\mu }$ is defined by [2]%
\begin{equation}
C_{\mu \nu }{}^{\lambda \sigma }=R_{\mu \nu }{}^{\lambda \sigma }-2R_{[\mu }%
\text{{}}^{[\lambda }g_{\nu ]}\text{{}}^{\sigma ]}+\frac{1}{3}Rg_{[\mu }%
\text{{}}^{\lambda }g_{\nu ]}\text{{}}^{\sigma },  \label{1}
\end{equation}%
with $R_{\mu \nu }{}^{\lambda \sigma },$ $R_{\mu \nu }$ and $R$ being,
respectively, the Riemann tensor, the Ricci tensor and the Ricci scalar of $%
\nabla _{\mu }$. The tensor $R_{\mu \nu \lambda \sigma }{}$ possesses
skewness in the indices of the pairs $\mu \nu $ and $\lambda \sigma ,$ but
the traditional index-pair symmetry ceases holding because of the
applicability of the cyclic identity%
\begin{equation}
^{\ast }R^{\lambda }{}_{\mu \nu \lambda }+2\nabla ^{\lambda }{}T_{\lambda
\mu \nu }^{\ast }+4{}T_{\mu }^{\ast }{}^{\lambda \tau }{}T_{\lambda \tau \nu
}=0,  \label{2}
\end{equation}%
where $T_{\mu \nu \lambda }{}{}$ is the torsion tensor of $\nabla _{\mu }$
which satisfies $T_{\mu \nu \lambda }{}{}=T_{[\mu \nu ]\lambda }{}{}$ by
definition, and the starred objects just come from adequate first left-right
dualizations [2,5]. As a consequence, $R_{\mu \nu }$ bears asymmetry while
the Bianchi identity reads%
\begin{equation}
\nabla ^{\rho }{}^{\ast }R_{\rho \mu \lambda \sigma }+2{}T{}{}_{\mu }^{\ast
}{}^{\rho \tau }R_{\rho \tau \lambda \sigma }=0.  \label{10}
\end{equation}%
Thus, $R_{\mu \nu \lambda \sigma }{}$ and $R_{\mu \nu }$ possess 36 and 16
independent components, respectively, whilst $C_{\mu \nu \lambda \sigma }{}$
possesses 20.

In either formalism, the gravitational curvature spinors of $\nabla _{\mu }$
are borne by the correspondence%
\begin{equation}
R_{\mu \nu \lambda \sigma }\leftrightarrow (\text{X}_{ABCD},\Xi _{A^{\prime
}B^{\prime }CD}),  \label{12}
\end{equation}%
which actually comprises the bivector constituents of the respective mixed
world-spin entity%
\begin{equation}
W_{\mu \nu (CD)}{}=\frac{1}{2}S_{CA^{\prime }}^{\lambda }{}S_{D}^{\sigma
A^{\prime }}R_{\mu \nu \lambda \sigma }{}.  \label{90}
\end{equation}%
The $S$-objects of Eq. (\ref{90}) stand for apposite connecting objects,
which must obey the "strong" relations (see Eq. (\ref{91}) below)%
\begin{equation}
S_{\mu A^{\prime }}^{(C}S_{\nu }^{D)A^{\prime }}=S_{A^{\prime }[\mu
}^{(C}S_{\nu ]}^{D)A^{\prime }}=S_{A^{\prime }[\mu }^{C}S_{\nu
]}^{DA^{\prime }}.  \label{26}
\end{equation}%
The X$\Xi $-entries thus have the symmetries%
\begin{equation}
\text{X}_{ABCD}=\text{X}_{(AB)(CD)},\text{ }\Xi _{A^{\prime }B^{\prime
}CD}=\Xi _{(A^{\prime }B^{\prime })(CD)}.  \label{13}
\end{equation}%
Since $R_{\mu \nu \lambda \sigma }\neq R_{\lambda \sigma \mu \nu }$, we also
have%
\begin{equation}
\text{X}_{ABCD}\neq \text{X}_{CDAB},\text{ }\Xi _{A^{\prime }B^{\prime
}CD}\neq \Xi _{CDA^{\prime }B^{\prime }},  \label{add1}
\end{equation}%
whence the X$\Xi $-spinors naively recover the number of degrees of freedom
of $R_{\mu \nu \lambda \sigma }$ as $18+18$. In the $\varepsilon $%
-formalism, one gets the expressions%
\begin{equation}
R_{AA^{\prime }BB^{\prime }CC^{\prime }DD^{\prime }}=\hspace{-1pt}%
(\varepsilon _{A^{\prime }B^{\prime }}\varepsilon _{C^{\prime }D^{\prime }}%
\text{X}_{ABCD}+\varepsilon _{AB}\varepsilon _{C^{\prime }D^{\prime }}\Xi
_{A^{\prime }B^{\prime }CD})+\text{c.c.}  \label{5}
\end{equation}%
and\footnote{%
The symbol "c.c." has been taken here as elsewhere to denote an overall
complex conjugate piece.}%
\begin{equation}
^{\ast }R_{AA^{\prime }BB^{\prime }CC^{\prime }DD^{\prime
}}=[(-i)(\varepsilon _{A^{\prime }B^{\prime }}\varepsilon _{C^{\prime
}D^{\prime }}\text{X}_{ABCD}-\varepsilon _{AB}\varepsilon _{C^{\prime
}D^{\prime }}\Xi _{A^{\prime }B^{\prime }CD})]+\text{c.c.},  \label{6}
\end{equation}%
with the $\varepsilon $-objects being the only covariant metric spinors for
the formalism allowed for. By definition, these latter objects enter the
equivalent metric correspondences\footnote{%
The uniqueness property of the $\varepsilon $-metric spinors is directly
related to their invariant spin-density character with respect to the action
of the Weyl gauge group [4]. The $\Sigma $-objects of Eqs. (\ref{91})
through (\ref{93}) amount to connecting objects for the $\varepsilon $%
-formalism.}%
\begin{equation}
2\Sigma _{AA^{\prime }(\mu }\Sigma _{\nu )B}{}^{A^{\prime }}=\varepsilon
_{AB}{}g_{\mu \nu },  \label{91}
\end{equation}%
\begin{equation}
g_{\mu \nu }{}=\Sigma _{\mu }^{AA^{\prime }}\Sigma _{\nu }^{BB^{\prime
}}\varepsilon _{AB}\varepsilon _{A^{\prime }B^{\prime }}  \label{92}
\end{equation}%
and%
\begin{equation}
\varepsilon _{AB}\varepsilon _{A^{\prime }B^{\prime }}=\Sigma _{AA^{\prime
}}^{\mu }\Sigma _{BB^{\prime }}^{\nu }g_{\mu \nu }.{}  \label{93}
\end{equation}

On the basis of the valence-reduction devices built up in Ref. [2], we can
expand the $\varepsilon $-formalism version of the X-spinor of (\ref{12}) as%
\begin{equation}
\text{X}_{ABCD}\hspace{-0.07cm}=\hspace{-0.07cm}\Psi _{ABCD}-\varepsilon
_{(A\mid (C}\xi _{D)\mid B)}-\frac{1}{3}\varkappa \varepsilon
_{A(C}\varepsilon _{D)B},  \label{9}
\end{equation}%
with the individual pieces [4]%
\begin{equation}
\hspace{-0.07cm}\Psi _{ABCD}=\text{X}_{(ABCD)}\hspace{-0.07cm},\text{ }\xi
_{AB}=\text{X}^{M}{}_{(AB)M},\text{ }\varkappa =\text{X}_{LM}{}^{LM}.
\label{7}
\end{equation}%
The $\Psi $-spinor of Eq. (\ref{9}) is usually taken as a typical wave
function for gravitons [1,9], and $\varkappa $ amounts to a complex-valued
world-spin invariant. In Ref. [5], $\xi _{AB}$ was supposed to account for a
wave function for dark matter. Its occurrence in the expansion (\ref{9}) is
essentially related to the appropriateness of the first of Eqs. (\ref{add1}%
). The complex valuedness of $\varkappa $ enables us to reinstate promptly
the number of complex independent components of X$_{ABCD}$ as $5+3+1$.
Hence, it may be said that the objects%
\begin{equation}
(\Psi _{ABCD},\xi _{AB},\varkappa ,\Xi _{A^{\prime }B^{\prime }CD})
\label{add2}
\end{equation}%
together determine both $R_{\mu \nu \lambda \sigma }$ and $^{\ast }R_{\mu
\nu \lambda \sigma }$ completely. It follows that we can spell out the
particular associations%
\begin{equation}
R_{\mu \nu }\leftrightarrow R_{AA^{\prime }BB^{\prime }}=\varepsilon
_{AB}\varepsilon _{A^{\prime }B^{\prime }}\func{Re}\varkappa -[(\varepsilon
_{A^{\prime }B^{\prime }}\xi _{AB}+\Xi _{A^{\prime }B^{\prime }AB})+\text{%
c.c.}]  \label{21}
\end{equation}%
and%
\begin{equation}
^{\ast }R^{\lambda }{}_{\mu \lambda \nu }\leftrightarrow R^{CC^{\prime
}}{}_{AA^{\prime }CC^{\prime }BB^{\prime }}=[i(\varepsilon _{A^{\prime
}B^{\prime }}\xi _{AB}-\frac{1}{2}\varepsilon _{AB}\varepsilon _{A^{\prime
}B^{\prime }}\varkappa -\Xi _{A^{\prime }B^{\prime }AB})]+\text{c.c.}.
\label{J}
\end{equation}%
Accordingly, the parts of $\varkappa $ are subject to [4]%
\begin{equation}
R=4\func{Re}\varkappa ,\text{ }^{\ast }R{}_{\mu \nu }{}^{\mu \nu }=4\func{Im}%
\varkappa ,  \label{23}
\end{equation}%
with Eq. (\ref{21}) thereby recovering the number of degrees of freedom of $%
R_{\mu \nu }$ as $1+6+9$.

\section{Spinor Transcription}

Equations (\ref{5}) and (\ref{9}) give right away the correspondence%
\footnote{%
Owing to the symmetry of the $\xi $-spinor, there is no need for staggering
its indices. Also, it should be stressed that $\varepsilon
_{(A}{}^{B}\varepsilon _{C)}{}^{D}=\varepsilon _{(A}{}^{(B}\varepsilon
_{C)}{}^{D)}$.}%
\begin{equation}
R_{\mu \nu }{}^{\lambda \sigma }\leftrightarrow \varepsilon ^{B^{\prime
}D^{\prime }}[\varepsilon _{A^{\prime }C^{\prime }}(\Psi
_{AC}{}^{BD}-\varepsilon _{(A}{}^{(B}\xi _{C)}^{D)}+\frac{1}{3}\varkappa
\varepsilon _{(A}{}^{B}\varepsilon _{C)}{}^{D})+\varepsilon _{AC}\Xi
_{A^{\prime }C^{\prime }}{}^{BD}]+\text{c.c.},  \label{c1}
\end{equation}%
which involves making some index substitutions in the expression (\ref{5})
so as to yield the pattern%
\begin{equation}
R_{\mu \nu }{}^{\lambda \sigma }\leftrightarrow R_{AA^{\prime }CC^{\prime
}}{}^{BB^{\prime }DD^{\prime }}.  \label{add3}
\end{equation}%
By implementing Eq. (\ref{21}) and invoking one of the relations (\ref{23}),
we obtain the irreducible relationship between $-2R_{[\mu }${}$^{[\lambda
}g_{\nu ]}${}$^{\sigma ]}$ and the configuration%
\begin{equation}
\varepsilon ^{B^{\prime }D^{\prime }}[\varepsilon _{A^{\prime }C^{\prime
}}(\varepsilon _{(A}{}^{(B}\xi _{C)}^{D)}-\varepsilon _{(A}{}^{B}\varepsilon
_{C)}{}^{D}\func{Re}\varkappa )-\frac{1}{2}\varepsilon _{AC}(\Xi _{A^{\prime
}C^{\prime }}{}^{BD}+\Xi ^{BD}{}_{A^{\prime }C^{\prime }})]+\text{c.c.}.
\label{c2}
\end{equation}%
It is worth mentioning that the derivation of the arrangement (\ref{c2})
takes up the complex conjugates of the blocks%
\begin{equation}
R_{(AC)}{}^{(B^{\prime }D^{\prime })}=-(\Xi _{AC}{}^{B^{\prime }D^{\prime
}}+\Xi ^{B^{\prime }D^{\prime }}{}_{AC})  \label{add5}
\end{equation}%
and%
\begin{equation}
R_{M}{}^{M}{}_{(A^{\prime }}{}^{(B^{\prime }}\varepsilon _{C^{\prime
})}{}^{D^{\prime })}=2(\varepsilon _{(A^{\prime }}{}^{B^{\prime
}}\varepsilon _{C^{\prime })}{}^{D^{\prime }}\func{Re}\varkappa -\varepsilon
_{(A^{\prime }}{}^{(B^{\prime }}\xi _{C^{\prime })}^{D^{\prime })}).
\label{add6}
\end{equation}%
For the $Rgg$-term of Eq. (\ref{1}), we get%
\begin{equation}
\frac{1}{3}Rg_{[\mu }\text{{}}^{\lambda }g_{\nu ]}\text{{}}^{\sigma
}\leftrightarrow \frac{2}{3}(\varepsilon _{A^{\prime }C^{\prime
}}\varepsilon ^{B^{\prime }D^{\prime }}\varepsilon _{(A}{}^{B}\varepsilon
_{C)}{}^{D}+\text{c.c.})\func{Re}\varkappa .  \label{c3}
\end{equation}

The combination of Eqs. (\ref{c1})-(\ref{c3}) with (\ref{1}) produces the
cancellation of the contributions that carry the $\varepsilon \xi $-pieces
and $\func{Re}\varkappa $. We are therefore led to the following expression
for $C_{\mu \nu }{}^{\lambda \sigma }$ 
\begin{equation}
\varepsilon ^{B^{\prime }D^{\prime }}[\varepsilon _{A^{\prime }C^{\prime
}}(\Psi _{AC}{}^{BD}+\frac{1}{3}\varepsilon _{(A}{}^{B}\varepsilon
_{C)}{}^{D}i\func{Im}\varkappa )+\frac{1}{2}\varepsilon _{AC}(\Xi
_{A^{\prime }C^{\prime }}{}^{BD}-\Xi ^{BD}{}_{A^{\prime }C^{\prime }})]+%
\text{c.c.},  \label{c4}
\end{equation}%
which transparently recovers the number of independent components of $C_{\mu
\nu }{}^{\lambda \sigma }$ as $10+1+9$, and thence $C_{\mu \nu \lambda
\sigma }$ gets completely determined out of%
\begin{equation}
(\Psi _{ABCD},\func{Im}\varkappa ,\Xi _{A^{\prime }B^{\prime }CD}{},\Xi
_{CDA^{\prime }B^{\prime }}{}).  \label{det1}
\end{equation}%
From (\ref{c4}), we see that $C_{\mu \nu }{}^{\lambda \sigma }\neq 0$
throughout spacetime since the overall terms%
\begin{equation}
\varepsilon _{A^{\prime }C^{\prime }}\varepsilon ^{B^{\prime }D^{\prime
}}(\Psi _{AC}{}^{BD}+\frac{1}{3}\varepsilon _{(A}{}^{B}\varepsilon
_{C)}{}^{D}i\func{Im}\varkappa ),\text{ }\frac{1}{2}\varepsilon
_{AC}\varepsilon ^{B^{\prime }D^{\prime }}(\Xi _{A^{\prime }C^{\prime
}}{}^{BD}-\Xi ^{BD}{}_{A^{\prime }C^{\prime }})  \label{c5}
\end{equation}%
as well as their conjugates are all nowhere-vanishing spinors in the
spacetime background we have been dealing with and, in addition, can not
cancel out one another. Yet, we may possibly have torsional spacetimes
wherein $\Psi _{ABCD}{}\equiv 0$, in which case the spinor expression for $%
C_{\mu \nu }{}^{\lambda \sigma }$ could be reset as%
\begin{equation}
\varepsilon ^{B^{\prime }D^{\prime }}[\frac{1}{3}(\varepsilon _{A^{\prime
}C^{\prime }}\varepsilon _{(A}{}^{B}\varepsilon _{C)}{}^{D}i\func{Im}%
\varkappa )+\frac{1}{2}\varepsilon _{AC}(\Xi _{A^{\prime }C^{\prime
}}{}^{BD}-\Xi ^{BD}{}_{A^{\prime }C^{\prime }})]+\text{c.c.}.  \label{add15}
\end{equation}%
Then, we have established that it is not possible to attain any general
condition for conformal flatness in a spacetime endowed with a torsionful
affinity even if the corresponding gravitational wave functions are taken to
vanish identically.

\section{Concluding Remarks}

In general relativity, the property $R_{\mu \nu \lambda \sigma }=R_{[\mu \nu
][\lambda \sigma ]}$ and the torsionless cyclic identity $^{\ast }R^{\lambda
}{}_{\mu \nu \lambda }=0$ entail imparting the index-pair symmetry to $%
R_{\mu \nu \lambda \sigma }{}$. Consequently, in the torsionless limiting
case, the curvature spinors of (\ref{12}) turn out to enjoy also the
symmetries X$_{ABCD}=$ X$_{CDAB}$ and $\Xi _{A^{\prime }B^{\prime }CD}=\Xi
_{CDA^{\prime }B^{\prime }}$, while $\func{Im}\varkappa =0$ by the second of
Eqs. (\ref{23}). Under this geometric circumstance, the $\Xi $-term of (\ref%
{c5}) vanishes identically and the expression (\ref{c4}) thus supplies us
with the correspondence [2,10]%
\begin{equation*}
C_{\mu \nu \lambda \sigma }\leftrightarrow \varepsilon _{A^{\prime
}B^{\prime }}\varepsilon _{C^{\prime }D^{\prime }}\Psi _{ABCD}{}+\text{c.c.},
\end{equation*}%
whence any generally relativistic Weyl tensor carries $10=20-1-9$ degrees of
freedom regardless of whether $R=0$ or not.\footnote{%
The expansion for an $\varepsilon $-formalism X-spinor as given in Refs.
[2,10] carries invariants $\Lambda $ and $\chi $ which obey the relations $%
\func{Re}\varkappa =2\chi =6\Lambda .$}

A spacetime that admits a Weyl tensor like the one expressed by (\ref{add15}%
) may, in a more or less conventional way, be designated as a conformally
flat spacetime. Nevertheless, its Weyl tensor must be defined locally as a
non-vanishing object, in contradistinction to the case of general
relativity. The word block "\textit{conformally flat Kopczy\'{n}ski solution}%
", referred to in the concluding Section of Ref. [5], may then look
legitimate.

We saw that $\xi _{AB}$ does not occur in the expression (\ref{c4}). Hence,
according to the picture proposed in Ref. [5], dark matter does not
contribute to any extent to the Weyl tensor of $\nabla _{\mu }.$ Neither do
it the cosmic microwave and dark energy backgrounds because these can never
be thought of as possessing a gravitational character [11].

Acknowledgement: I should acknowledge a reviewer for a suggestion that has
produced an enrichment of Section 2.

\end{document}